\documentclass[10pt]{blois07}
\usepackage{graphicx,amsmath}
\usepackage{cite,./mcite}

\begin{document}

\newcommand{\orders}[1]{\ensuremath{\mathcal{O}\left(s^{#1}\right)}}
\newcommand{\ordershat}[1]{\mathcal{O}\left(\hat{s}^{#1}\right)}
\newcommand{\orderslambda}[1]{\mathcal{O}\left(s_\Lambda^{#1}\right)}
\newcommand{\ordereps}[1]{\ensuremath{\mathcal{O}\left(\epsilon^{#1}\right)}}
\newcommand{\order}[1]{\mathcal{O}\left(#1\right)}
\newcommand{\dk}{d^{D-2}{\bf k}}
\newcommand{\dka}{d^{D-2}{{\bf k}_a}}\newcommand{\dkb}{d^{D-2}{{\bf k}_b}}
\newcommand{\dkl}{d^{D-2}{{\bf k}_l}}\newcommand{\dki}{d^{D-2}{{\bf k}_i}}
\newcommand{\dkone}{d^{D-2}{{\bf k}_1}}
\newcommand{\dktwo}{d^{D-2}{{\bf k}_2}}
\newcommand{\dktwoeps}{\frac{d^{D-2}{{\bf k}_2}}{\mu^{2\epsilon}(2\pi)^{D-4}}}
\newcommand{\dkjet}{d^{D-2}{{\bf k}_J}}
\newcommand{\dkjetone}{d^{D-2}{{\bf k}_{J_1}}}
\newcommand{\dkjettwo}{d^{D-2}{{\bf k}_{J_2}}}
\newcommand{\dkpure}{d^{D-2}{{\bf k}}}
\newcommand{\dkprime}{d^{D-2}{{\bf k}'}}
\newcommand{\dlambdeps}{\frac{d^{D-2}{{\bf \Lambda}}}{\mu^{2\epsilon}(2\pi)^{D-4}}}
\newcommand{\ki}{{\bf k}_i}
\newcommand{\kim}{{\bf k}_{i-1}}
\newcommand{\kip}{{\bf k}_{i+1}}
\newcommand{\kj}{{\bf k}_j}
\newcommand{\kjm}{{\bf k}_{j-1}}\newcommand{\kjp}{{\bf k}_{j+1}}
\newcommand{\km}{{\bf k}_m}
\newcommand{\kmm}{{\bf k}_{m-1}}
\newcommand{\kmp}{{\bf k}_{m+1}}
\newcommand{\kl}{{\bf k}_l}
\newcommand{\klm}{{\bf k}_{l-1}}
\newcommand{\klp}{{\bf k}_{l+1}}
\newcommand{\kn}{{\bf k}_n}
\newcommand{\knm}{{\bf k}_{n-1}}
\newcommand{\knp}{{\bf k}_{n+1}}
\newcommand{\ka}{{\bf k}_a}
\newcommand{\kb}{{\bf k}_b}
\newcommand{\kone}{{\bf k}_1}
\newcommand{\ktwo}{{\bf k}_2}
\newcommand{\kjet}{{\bf k}_J}
\newcommand{\kjetone}{{\bf k}_{J_1}}
\newcommand{\kjettwo}{{\bf k}_{J_2}}
\newcommand{\kpure}{{\bf k}}
\newcommand{\kprime}{{\bf k}'}
\newcommand{\oma}{\omega_0({\bf k}_a)}
\newcommand{\om}[1]{\omega_0(#1)}
\newcommand{\omall}{\omega_0^{LL}({\bf k}_a)}
\newcommand{\omll}[1]{\omega_0^{LL}(#1)}
\newcommand{\del}[1]{\delta^{(2)}\left(#1\right)}
\newcommand{\non}{\nonumber\\}
\newcommand{\asquare}[4]{\left|\mathcal{A}(#1,#2,#3,#4)\right|^2}
\newcommand{\bssquare}[4]{\left|\mathcal{B}_s(#1,#2,#3,#4)\right|^2}
\newcommand{\btssquare}[4]{\left|\widetilde\mathcal{B}_s(#1,#2,#3,#4)\right|^2}
\newcommand{\bsquare}[4]{\left|\mathcal{B}(#1,#2,#3,#4)\right|^2}
\newcommand{\agsquare}[4]{\left|\mathcal{A}_{2g}(#1,#2,#3,#4)\right|^2}
\newcommand{\aqsquare}[4]{\left|\mathcal{A}_{2q}(#1,#2,#3,#4)\right|^2}
\newcommand{\asbar}{\bar\alpha_s}
\newcommand{\seplog}[2]{\ln\frac{s_\Lambda}{\sqrt{#1^2 #2^2}}}
\newcommand{\qi}{{\bf q}_i}\newcommand{\qim}{{\bf q}_{i-1}}\newcommand{\qip}{{\bf q}_{i+1}}
\newcommand{\qj}{{\bf q}_j}\newcommand{\qjm}{{\bf q}_{j-1}}\newcommand{\qjp}{{\bf q}_{j+1}}
\newcommand{\ql}{{\bf q}_l}\newcommand{\qlm}{{\bf q}_{l-1}}\newcommand{\qlp}{{\bf q}_{l+1}}
\newcommand{\qn}{{\bf q}_n}\newcommand{\qnm}{{\bf q}_{n-1}}\newcommand{\qnp}{{\bf q}_{n+1}}\newcommand{\qnpp}{{\bf q}_{n+2}}
\newcommand{\qone}{{\bf q}_1}\newcommand{\qtwo}{{\bf q}_2}
\newcommand{\qa}{{\bf q}_a}\newcommand{\qb}{{\bf q}_b}
\newcommand{\qt}{\tilde{\bf q}}
\newcommand{\dqa}{d^{D-2}{{\bf q}_a}\;}\newcommand{\dqb}{d^{D-2}{{\bf q}_b}\;}
\newcommand{\dqi}{d^{D-2}{{\bf q}_i}\;}
\newcommand{\dqt}{d^{D-2}{\tilde{\bf q}}\;}
\newcommand{\omhat}{\hat{\omega}_0}\newcommand{\omhati}{\hat{\omega}_i}\newcommand{\omhatim}{\hat{\omega}_{i-1}}\newcommand{\omhatl}{\hat{\omega}_l}\newcommand{\omhatn}{\hat{\omega}_n}\newcommand{\omhatlm}{\hat{\omega}_{l-1}}

\newcommand{\shat}{\hat{s}}\newcommand{\that}{\hat{t}}\newcommand{\uhat}{\hat{u}}
\newcommand{\lambd}{{\bf{\Lambda}}}\newcommand{\delt}{{\bf{\Delta}}}

\title{NLO jet production in $k_T$ factorization}
\author{J. Bartels$^1$, A. Sabio Vera$^2$, F. Schwennsen$^1$\protect\footnote{ ~ talk presented at EDS07}
}
\institute{$^1$II. Institut f\"ur Theoretische Physik, Universit\"at Hamburg, Luruper Chaussee 149, D-22761~Hamburg, Germany, \\
$^2$Physics Department, Theory Division, CERN, CH-1211 Geneva 23, Switzerland
}
\maketitle
\begin{abstract}
We discuss the inclusive production of jets in the central region of rapidity in the context of $k_T$--factorization at next--to--leading order (NLO). Calculations are performed in the Regge limit making use of the NLO BFKL results. We introduce a jet cone definition and carry out a proper phase--space separation into multi--Regge and quasi--multi--Regge kinematic regions. We discuss two situations: scattering of highly virtual photons, which requires a symmetric energy scale to separate impact factors from the gluon Green's function, and hadron--hadron collisions, where a non--symmetric scale choice is needed.
\end{abstract}

\section{Introduction}

An accurate  knowledge of perturbative QCD is an essential ingredient in phenomenological studies at present and future colliders. At high center of mass energies the theoretical study of multijet events becomes an increasingly important task. In the context of collinear factorization the calculation of multijet production is complicated because of the large number of contributing diagrams. There is, however, a region of phase space where it is indeed possible to describe the production of a large number of jets: the Regge asymptotics (small--$x$ region) of scattering amplitudes. If the jets are well separated in rapidity, the corresponding matrix element factorizes with effective vertices for the jet production connected by a chain of $t$-channel Reggeons.  A perturbative analysis of these diagrams shows that it is  possible to resum contributions of the form $\left(\alpha_s \ln{s}\right)^n$  to all orders, with $\alpha_s$ being the coupling constant for the strong  interaction. This can be achieved by means of the  Balitsky--Fadin--Kuraev--Lipatov (BFKL) equation~\cite{BFKL}.

The BFKL approach relies on the concept of the {\it Reggeized gluon} or {\it Reggeon}. In Regge asymptotics colour octet 
exchange can be effectively described by a $t$--channel gluon with its propagator being modified by a multiplicative factor depending on a power of $s$. 
This power corresponds to the {\it gluon Regge trajectory} which is a function of the transverse momenta and is divergent in the infrared. This divergence is removed when real emissions are included using gauge invariant Reggeon--Reggeon--gluon couplings. This allows us to describe scattering amplitudes with a large number of partons in the final state. The $\left(\alpha_s \ln{s}\right)^n$ terms correspond to the leading--order (LO) approximation and provide a simple picture of the underlying physics. This approximation has limitations: in leading order both $\alpha_s$ and the factor scaling the energy $s$ in the resummed logarithms, $s_0$, are free parameters not determined by the theory. These free parameters can be fixed if next--to--leading terms $\alpha_s \left(\alpha_s \ln{s}\right)^n$ are included~\cite{FC}. At this improved accuracy, diagrams contributing to the running of the coupling have to be included, and also $s_0$ is not longer undetermined. The phenomenological importance of the NLO effects has been recently shown in the scattering of virtual photons into vector mesons \cite{Ivanov}
as well as for azimuthal angle decorrelations in Mueller--Navelet jets in Refs.~\cite{VV} and in Deep Inelastic Scattering in Ref.~\cite{Vera:2007dr}.

While at LO the only emission vertex -- the Reggeon-Reggeon-gluon vertex -- can be identified with the production of one jet, at NLO also Reggeon-Reggeon-gluon-gluon and Reggeon-Reggeon-quark-antiquark vertices enter the game. 
In this contribution we are interested in the description of the inclusive production of a single jet in the NLO BFKL formalism. The relevant events will be those with only one jet produced in the central rapidity region of the detector.
Due to these new emission vertices at NLO we have to introduce a jet definition discriminating between the production of one or two jets by two particles. It is not sufficient to simply start from the fully integrated emission vertex available in the literature \cite{FC}. Rather, we have to carefully separate all the different contributions in its unintegrated form before we can combine them. 

The present text is focused on describing the main elements of the analysis presented in Refs.~\cite{BS}. There we show that this procedure enables us to determine the right choice of energy scales relevant for the process. Particular attention is given to the separation of multi-Regge and quasi-multi-Regge kinematics. There we also discuss similarities and discrepancies with the 
earlier work of Ref.~\cite{Ostrovsky:1999kj}.

As it turns out, the scale of the two projectiles in the scattering process has a large impact on the structure of the result. The jet vertex can not be constructed without properly defining the interface to the scattering objects. To show this, we will perform this study for two different cases: the jet production in the scattering of two photons with large and similar virtualities, and in hadron-hadron collisions. In the former case the cross section has a factorized form in terms of the photon impact factors and of the  gluon Green's function which is valid in the Regge limit. In the latter case, since the momentum scale of the hadron is substantially lower than the typical $k_T$ entering the production vertex, the gluon Green's function for hadron-hadron collisions has a slightly different BFKL kernel which, in particular, also incorporates some $k_T$-evolution from the nonperturbative, and model dependent, proton impact factor to the perturbative jet production  vertex. 

Our final expression for the cross section of the jet production in hadron-hadron scattering contains an \emph{unintegrated gluon density}. This density depends on the longitudinal momentum fraction -- as it also happens in the  conventional collinear factorization -- and on the transverse momentum $k_T$. 
As an important ingredient, the hard subprocesse (in our case, the production vertex) depends on the (off-shell) initial parton momenta.
This scheme of $k_T$--factorization has been introduced by Catani et al. \cite{Catani:1990eg}, and up to now it has been considered only at LO.
Our results, valid in the small--$x$ limit, show that it is possible to
extend this scheme to NLO.


\section{Inclusive jet production at LO}

Let us start  the discussion by considering the interaction between two photons 
with large virtualities $Q_{1,2}^2$ in the Regge limit 
$s \gg |t|\sim Q_1^2\sim Q_2^2$. In this region the total cross section can be 
written as a convolution of the photon impact factors with the gluon Green's 
function, {\it i.e.}
\begin{equation}
  \label{eq:total}
\sigma(s) = \int\frac{d^2 {\bf k}_a}{2\pi\ka^2}\int\frac{d^2{\bf k}_b}{2\pi\kb^2} \, 
\Phi_A(\ka) \, \Phi_B(\kb) \,\int_{\delta-i\infty}^{\delta+i\infty}\frac{d\omega}{2\pi i} \left(\frac{s}{s_0}\right)^\omega f_\omega(\ka,\kb).
\end{equation}
A convenient choice for the energy scale is $s_0=|\ka|\,|\kb|$ which  
naturally introduces the rapidities $y_{\tilde A}$ and $y_{\tilde B}$ of the 
emitted particles with momenta $p_{\tilde A}$ and $p_{\tilde B}$ given that 
$\left(\frac{s}{s_0}\right)^\omega = e^{\omega(y_{\tilde A}-y_{\tilde B})}$.

The gluon Green's function $f_\omega$ corresponds to the solution of the BFKL 
equation
\begin{eqnarray}
  \label{eq:bfklequation}
  \omega f_\omega(\ka,\kb) &=& \delta^{(2)}(\ka-\kb)+\int d^2{\bf k}\;\mathcal{K}(\ka,\kpure)f_\omega(\kpure,\kb),
\end{eqnarray}
with kernel
\begin{eqnarray}
  \mathcal{K}(\ka,\kpure) &=& 2 \, \omega(\ka^2) \, 
\delta^{2}(\ka-\kpure) + \mathcal{K}_r(\ka,\kpure),
\end{eqnarray}
where $\omega (\ka^2)$ is the gluon Regge trajectory and $\mathcal{K}_r$ is the real 
emission contribution to the kernel which is of special interest in the following.

\begin{center}
\begin{figure}[htbp]
  \centering
  \includegraphics[height=5cm]{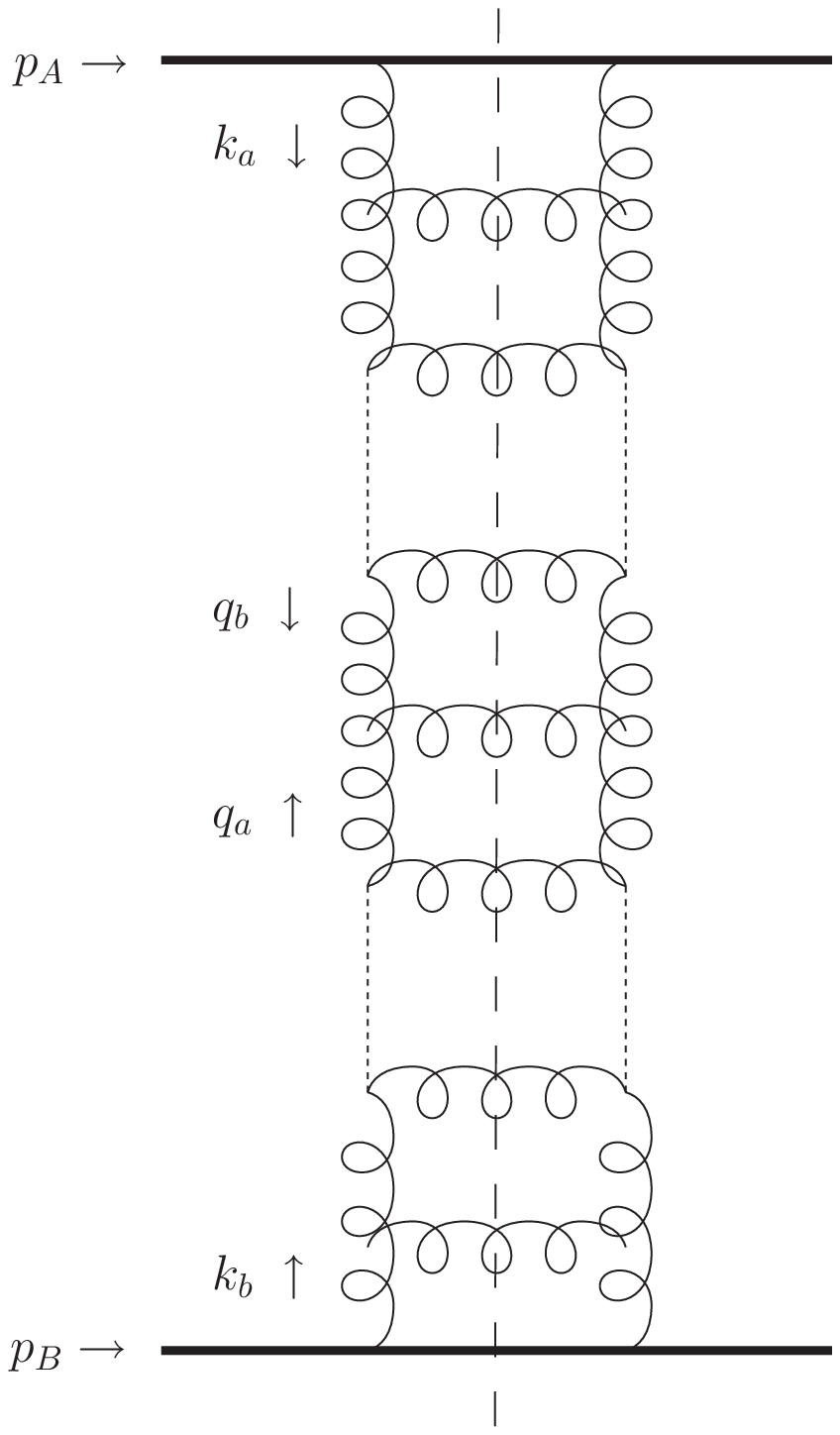}
  \hspace{2cm}
  \includegraphics[height=5cm]{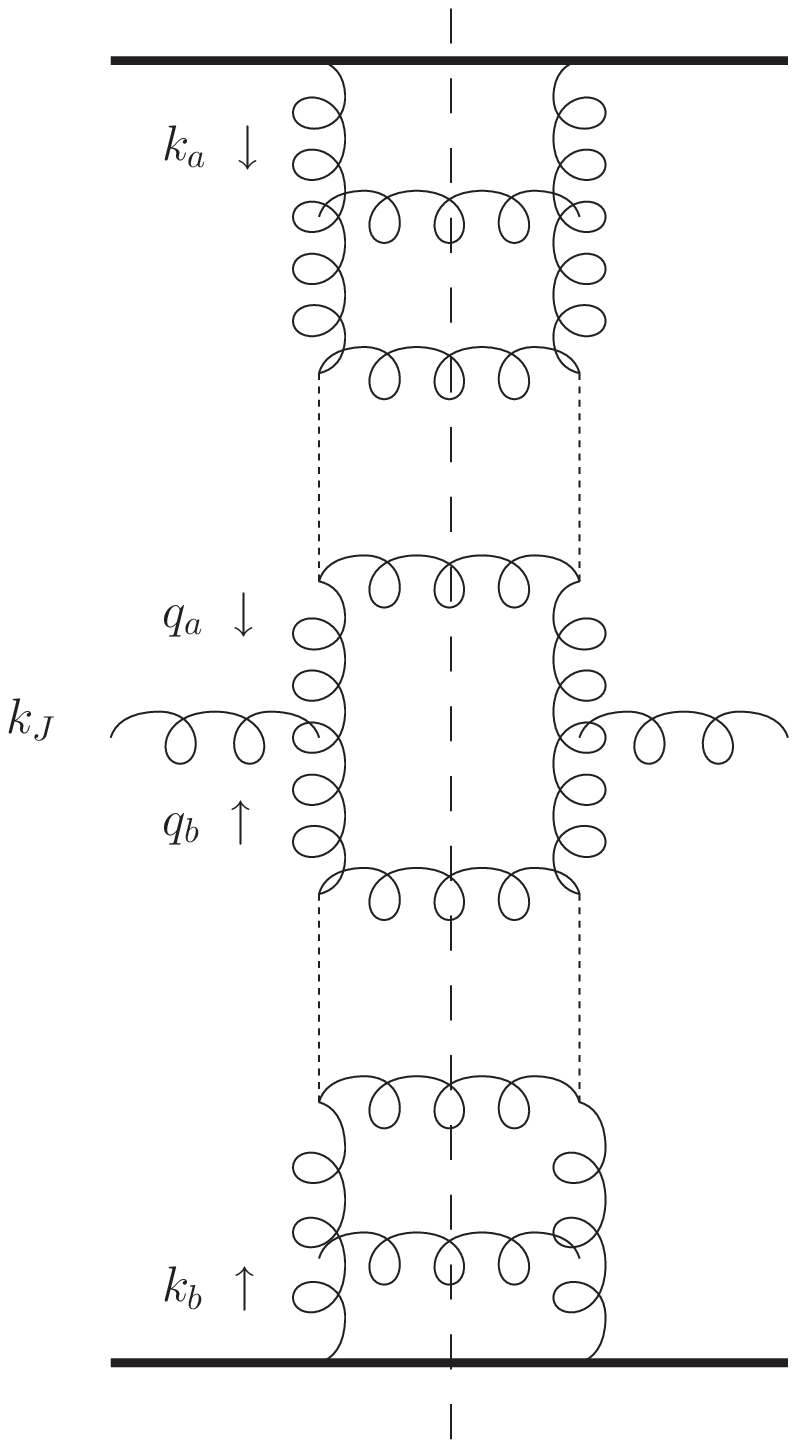}
  \caption{Total cross section and inclusive one jet production in the BFKL 
approach.}
  \label{fig:crosslo}
\end{figure}
\end{center}

It is possible to single out one gluon emission by extracting its  
emission probability from the BFKL kernel. 
By selecting one emission to be exclusive we factorize the gluon Green's 
function into two components. Each of them connects one of the external 
particles to the jet vertex, and depends on the total energies of the 
subsystems $s_{AJ} = (p_A+q_b)^2$ and $s_{BJ} = (p_B+q_a)^2$, respectively. 
We have drawn a graph indicating this separation in Fig.~\ref{fig:crosslo}. 
The symmetric situation suggests the choices $s_0^{(AJ)} =|\ka|\,|\kjet|$ and 
$s_0^{(BJ)} = |\kjet|\,|\kb|$, respectively, as the suitable energy scales for 
the subsystems. These choices can be related to the relative rapidity between 
the jet and the external particles. To set the ground for the NLO discussion 
of next section we introduce an additional integration over the rapidity 
$\eta$ of the central system in the form
\begin{multline}
\frac{d\sigma}{d^{2}\kjet dy_J} =
\int d^2 \qa \int d^2 \qb \int d\eta
\left[\int\frac{d^2 \ka}{2\pi\ka^2} \, \Phi_A(\ka) \, \int_{\delta-i\infty}^{\delta+i\infty}\frac{d\omega}{2\pi i} e^{\omega(y_A-\eta)} f_\omega(\ka,\qa)\right]\\
\times \mathcal{V}(\qa,\qb,\eta;\kjet,y_J)\;
\times \left[\int\frac{d^2 \kb}{2\pi\kb^2} \, 
\Phi_B(\kb) \int_{\delta-i\infty}^{\delta+i\infty}\frac{d\omega'}{2\pi i} e^{\omega'(\eta-y_B)}f_{\omega'}(-\qb,-\kb)\right] 
\label{eq:masterformula1}
\end{multline}
with the LO emission vertex being
\begin{equation}
\mathcal{V}(\qa,\qb,\eta;\kjet,y_J) = \mathcal{K}_r^{\rm (Born)}\left(\qa,-\qb\right) \, \del{\qa+\qb-\kjet}\,\delta(\eta-y_J). 
\label{eq:jetvertexloy}
\end{equation}

In hadron--hadron collisions the colliding external particles do not provide a perturbative scale.
There the jet is the only hard scale 
in the process and we have to deal with an asymmetric situation. In such a 
configuration the 
scales $s_0$ should be chosen as $\kjet^2$ alone. At LO accuracy $s_0$ is 
arbitrary and we are indeed free to make this choice. At this stage it is 
possible to introduce the concept of {\it unintegrated gluon 
density} in the hadron. This represents the probability of resolving a gluon 
carrying a longitudinal momentum fraction $x$ from the incoming hadron, and 
with a certain transverse momentum $k_T$. Its relation to the gluon Green's 
function would be
\begin{equation}
  \label{eq:updflo}
  g(x,\kpure) = \int\frac{d^2 {\bf q}}{2\pi {\bf q}^2}\,\Phi_{P}({\bf q})\,\int_{\delta-i\infty}^{\delta+i\infty}\frac{d\omega}{2\pi i}\, x^{-\omega} f_\omega({\bf q},\kpure).
\end{equation}
With this new interpretation we can then rewrite Eq.~\eqref{eq:masterformula1} 
as
\begin{multline}
\label{eq:masterformula2}
\frac{d\sigma}{d^{2}\kjet dy_J} = \int d^2 \qa\int dx_1 \int d^2 \qb\int  dx_2\;g(x_1,\qa)g(x_2,\qb)\mathcal{V}(\qa,x_1,\qb,x_2;\kjet,y_J),
\end{multline}
with the LO jet vertex for the asymmetric situation being
\begin{multline}
\label{eq:jetvertexlo}
\mathcal{V}(\qa,x_1,\qb,x_2;\kjet,y_J)=\mathcal{K}_r^{\rm (Born)}\left(\qa,-\qb\right)\\
\times \del{\qa+\qb-\kjet}\,\delta\left(x_1-\sqrt{\frac{\kjet^2}{s}}e^{y_J}\right)\delta\left(x_2-\sqrt{\frac{\kjet^2}{s}}e^{-y_J}\right).
\end{multline}

\section{Inclusive jet production at NLO}

It is possible to follow a similar approach when jet 
production is considered at NLO. The crucial step in this direction is 
to modify  the LO jet vertex of Eq.~\eqref{eq:jetvertexloy} and 
Eq.~\eqref{eq:jetvertexlo} to include new configurations present 
at NLO. We show how this is done in the following first subsection. 
In the second subsection we implement this vertex in a scattering process.
In case of hadron--hadron scattering we extend the concept of unintegrated gluon 
density of Eq.~\eqref{eq:updflo} to NLO accuracy. Most importantly, it is 
shown that a correct choice of intermediate energy scales in this case 
implies a modification of the impact factors, the jet vertex, and the evolution kernel.

\subsection{The NLO jet vertex}

For those 
parts of the NLO kernel responsible for one gluon production we proceed in exactly the 
same way as at LO. The treatment of those terms related to two particle production is more complicated since  
for them it is necessary to introduce a jet algorithm. In general terms, if the two 
emissions generated by the kernel are nearby in phase space they will be considered as 
one single jet, otherwise one of them will be identified as the jet whereas the other 
will be absorbed as an untagged inclusive contribution. Hadronization effects in the 
final state are neglected and we simply define a cone of radius $R_0$ in the 
rapidity--azimuthal angle space such that two particles form a single jet if 
$R_{12} \equiv \sqrt{(\phi_1-\phi_2)^2+(y_1-y_2)^2} < R_0$. As long as only 
two emissions are involved this is equivalent to the $k_T$--clustering algorithm. 

To introduce the jet definition in the $2 \rightarrow 2$ components of the 
kernel it is convenient to combine the gluon and quark matrix 
elements together with the MRK contribution:
\begin{align}
&\left({\cal K}_{Q\bar Q}+ {\cal K}_{GG}  \right) (\qa,-\qb) \equiv 
\int\dktwo\int dy_2\,\bsquare{\qa}{\qb}{\kone}{\ktwo} \nonumber\\
 =& \int\dktwo\int dy_2\,
\Bigg\{\aqsquare{\qa}{\qb}{\kone}{\ktwo}+\agsquare{\qa}{\qb}{\kone}{\ktwo}\theta(s_\Lambda-s_{12})\nonumber\\
&-{\cal K}^{\rm (Born)}(\qa,\qa-\kone) \, {\cal K}^{\rm (Born)}(\qa-\kone,-\qb)\;\frac{1}{2}\,\theta\left(\ln\frac{s_{\Lambda}}{\ktwo^2}-y_2\right)\theta\left(y_2-\ln\frac{\kone^2}{s_{\Lambda}}\right) \Bigg\}, 
\label{eq:defbsquare}
\end{align}
with ${\cal A}_{2P}$ being the two particle production amplitudes. At NLO it is necessary to separate multi-Regge kinematics (MRK) from quasi-multi-Regge kinematics (QMRK) in a distinct way. With this purpose we introduce an additional scale, $s_\Lambda$. The meaning of MRK is that the invariant mass of two emissions is considered larger than $s_\Lambda$ while in QMRK the invariant mass of one pair of these emissions is below this scale.

The NLO version of Eq.~\eqref{eq:jetvertexloy} then reads
\begin{align}
\mathcal{V}(\qa,\qb,\eta;\kjet,y_J)= & \left(\mathcal{K}_r^{\rm (Born)} +\mathcal{K}_r^{\rm (virtual)}\right)(\qa,-\qb) \Big|_{(a)}^{[y]}\non
  &\hspace{-2cm}+ \int\dktwo\;dy_2\bsquare{\qa}{\qb}{\kjet-\ktwo}{\ktwo}\theta(R_0-R_{12})\Big|_{(b)}^{[y]}\non
  &\hspace{-2cm}+ 2\int\dktwo\;dy_2\bsquare{\qa}{\qb}{\kjet}{\ktwo}\theta(R_{J2}-R_0)\Big|_{(c)}^{[y]}.\label{eq:jetvertexnloy}
\end{align}
In this expression we have introduced the notation 
\begin{align}
  &\Big|_{(a,b)}^{[y]}&&\hspace{-2.6cm}=\del{\qa+\qb-\kjet} \delta (\eta - y^{(a,b)}),  \\
  &\Big|_{(c)}^{[y]}&&\hspace{-2.6cm}=\del{\qa+\qb-\kjet-\ktwo} \delta \left(\eta-y^{(c)}\right). 
\end{align}

The various jet configurations demand several $y$ and $x$ configurations. These are related to the properties of the produced jet in different ways 
depending on the origin of the jet: if only one gluon was produced in MRK this 
corresponds to the configuration (a) in the table below, if two particles in 
QMRK form a jet then we have the case (b), and finally case (c) if the jet is 
produced out of one of the partons in QMRK. The factor of 2 in the last term 
of Eq.~\eqref{eq:jetvertexnloy} accounts for the possibility that either 
emitted particle can form the jet. The vertex can be written in a similar 
way if one chooses to work in $x$ configuration language.
Just by kinematics we get the explicit expressions for the different $x$ configurations listed in the following table:
\begin{center}
\begin{tabular}[h!]{c|c|cc}
  JET  & $y$ configurations & \multicolumn{2}{c}{$x$ configurations}
\\\hline
a) \raisebox{-1ex}{\includegraphics[height=.7cm]{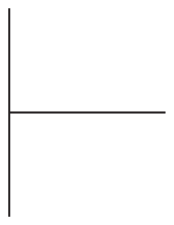}}  & 
   $y^{(a)}=y_J$ & $x_1^{(a)}=\frac{|\kjet|}{\sqrt{s}}e^{y_J}$ & 
   $x_2^{(a)}=\frac{|\kjet|}{\sqrt{s}}e^{-y_J}$ \\
b) \raisebox{-1.5ex}{\includegraphics[height=.7cm]{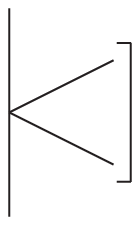}} & 
   $y^{(b)}=y_J$ & $x_1^{(b)}=\frac{\sqrt{\Sigma}}{\sqrt{s}}e^{y_J}$ & 
   $x_2^{(b)}=\frac{\sqrt{\Sigma}}{\sqrt{s}}e^{-y_J}$\\
c) \raisebox{-2ex}{\includegraphics[height=.7cm]{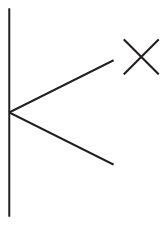}} & 
   $y^{(c)}=\frac{1}{2}\ln\frac{x_1^{(c)}}{x_2^{(c)}}$ & $ x_1^{(c)}=\frac{|\kjet|}{\sqrt{s}}e^{y_J}+\frac{|\ktwo|}{\sqrt{s}}e^{y_2}$ &   {\small $ x_2^{(c)}=\frac{|\kjet|}{\sqrt{s}}e^{-y_J}+\frac{|\ktwo|}{\sqrt{s}}e^{-y_2}$}
\end{tabular}
\end{center}


The NLO virtual correction to the one--gluon emission kernel, 
${\cal K}^{(v)}$, was originally calculated in 
Ref.~\cite{FFF}. It includes explicit infrared divergences which are canceled by the real contributions. 
The introduction of the jet definition divides the phase space into different sectors.
Only if the divergent terms belong to the same 
configuration this cancellation can be shown analytically. 
With this in mind we 
add the singular parts of the two particle production 
$|\mathcal{B}_s|^2$ in the configuration $(a)$ 
multiplied by $0=1-\theta(R_0-R_{12})-\theta(R_{12}-R_0)$:
\begin{multline}
\mathcal{V} = \bigg[\left(\mathcal{K}_r^{\rm (Born)}+\mathcal{K}_r^{\rm (virtual)}\right)(\qa,-\qb)+\int\dktwo\;dy_2\bssquare{\qa}{\qb}{\kjet-\ktwo}{\ktwo}\bigg] \Big|_{(a)}\\
+ \int\dktwo\;dy_2\bigg[\bsquare{\qa}{\qb}{\kjet-\ktwo}{\ktwo}\Big|_{(b)}
-\bssquare{\qa}{\qb}{\kjet-\ktwo}{\ktwo}\Big|_{(a)}\bigg]\\
\times\theta(R_0-R_{12})+2\int\dktwo\;dy_2\bigg[\bsquare{\qa}{\qb}{\kjet}{\ktwo}\theta(R_{J2}-R_0)\Big|_{(c)}\\
-\bssquare{\qa}{\qb}{\kjet-\ktwo}{\ktwo}\theta(R_{12}-R_0)\theta(|\kone|-|\ktwo|)\Big|_{(a)}\bigg].
\label{eq:freeofsing}
\end{multline}

The cancellation of divergences within the first line is now the same 
as in the calculation of the full NLO kernel.
The remainder is explicitly free 
of divergences as well since these have been subtracted out.

\subsection{Embedding of the jet vertex}

The NLO corrections to the kernel have been derived in the situation of the scattering of two objects with an intrinsic hard scale. Hence in the case of $\gamma^*\gamma^*$ scattering the equation \eqref{eq:masterformula1} is valid also at NLO if we replace the building blocks by their NLO counterparts. The most important piece being the jet vertex, which should be replaced by the one derived in the previous subsection.

We now turn to the case of hadron collisions where 
MRK has to be necessarily modified to include some evolution in the 
transverse momenta, since the momentum of the jet will be much 
larger than the typical transverse scale associated to the hadron.
In the LO case we have already explained that, in order to move from the symmetric 
case to the asymmetric one, it is needed to change the energy scale.
The independence of the result from this choice is guaranteed by a 
compensating modification of the impact factors
\begin{eqnarray}
\label{newimpactfactor}
 \widetilde{\Phi}(\ka)&=& \Phi(\ka) -\frac{1}{2}{\ka^2}\int d^2 {\bf q} 
\frac{\Phi^{\rm (Born)}({\bf q})}{{\bf q}^2}\mathcal{K}^{\rm (Born)}({\bf q},\ka)
\ln\frac{{\bf q}^2}{\ka^2}
\end{eqnarray}
and the evolution kernel
\begin{eqnarray}
\label{newkernel}
  \widetilde{\mathcal{K}}(\qone,\qtwo) &=& \mathcal{K}(\qone,\qtwo)
-\frac{1}{2}\int d^2 {\bf q} \, \mathcal{K}^{\rm (Born)}(\qone,{\bf q}) 
\, \mathcal{K}^{\rm (Born)}({\bf q},\qtwo)\ln\frac{{\bf q}^2}{\qtwo^2},
\end{eqnarray}
which corresponds to the first NLO term of a collinear 
resummation \cite{Salam}.

The emission vertex couples as a kind of impact factor to both  Green's functions and receives two such modifications:
\begin{eqnarray}
\label{newemissionvertex}
  \widetilde{\cal V}(\qa,\qb) &=& {\cal V}(\qa,\qb)
-\frac{1}{2}\int d^2 {\bf q} \,  
\mathcal{K}^{\rm (Born)}(\qa,{\bf q}) {\cal V}^{\rm (Born)}({\bf q},\qb)
\ln\frac{{\bf q}^2}{({\bf q}-\qb)^2}\nonumber\\
&&\hspace{1cm}-\frac{1}{2}\int d^2 {\bf q} \, {\cal V}^{\rm (Born)}(\qa,{\bf q}) \,
\mathcal{K}^{\rm (Born)}({\bf q},\qb)\ln\frac{{\bf q}^2}{(\qa-{\bf q})^2}.
\end{eqnarray}

\section{Conclusions}

We have extended the NLO BFKL calculations to derive a NLO jet
production vertex in $k_T$--factorization. 
Our procedure was to `open' the BFKL kernel to introduce a jet definition at NLO in a consistent way.  
As the central result, we have defined the off-shell jet production 
vertex and have shown how it can be used in the context of $\gamma^*\gamma^*$ 
or of hadron--hadron scattering to calculate inclusive single jet cross sections. 
For this purpose we have formulated, on the basis of the NLO BFKL equation, a
NLO unintegrated gluon density valid in the small--$x$ regime.
More recently, a slightly
different $k_T$--factorization scheme has been investigated \cite{Collins:2007ph}. A precise
analysis of the connection between the two approaches is in progress.

\noindent
{\bf Acknowledgments:} 
F.S. is supported by the Graduiertenkolleg ``Zuk\"unftige Entwicklungen in der Teilchenphysik''. Discussions with V.~Fadin and L.~Lipatov are gratefully acknowledged.  

\begin{footnotesize}
\bibliographystyle{blois07} 
{\raggedright

\providecommand{\etal}{et al.\xspace}
\providecommand{\href}[2]{#2}
\providecommand{\coll}{Coll.}
\catcode`\@=11
\def\@bibitem#1{%
\ifmc@bstsupport
  \mc@iftail{#1}%
    {;\newline\ignorespaces}%
    {\ifmc@first\else.\fi\orig@bibitem{#1}}
  \mc@firstfalse
\else
  \mc@iftail{#1}%
    {\ignorespaces}%
    {\orig@bibitem{#1}}%
\fi}%
\catcode`\@=12

}
\end{footnotesize}
\end{document}